# xQSM: Quantitative Susceptibility Mapping with Octave Convolutional and Noise Regularized Neural Networks

Yang Gao[1], Xuanyu Zhu[1], Bradford A. Moffat[2], Rebecca Glarin[2,3], Alan H. Wilman[4], G. Bruce Pike[5], Stuart Crozier[1], Feng Liu[1], Hongfu Sun[1]*

[1]School of Information Technology and Electrical Engineering, University of Queensland, Brisbane, Australia

[2]Melbourne Brain Centre Imaging Unit, Department of Medicine and Radiology, The University of Melbourne, Parkville, Australia

[3]Department of Radiology, Royal Melbourne Hospital, Parkville, Australia

[4]Department of Biomedical Engineering, University of Alberta, Edmonton, Canada

[5]Departments of Radiology and Clinical Neurosciences, University of Calgary, Calgary, Canada

**Correspondence**: Hongfu Sun

**Address**: Room 538, General Purpose South (Building 78), University of Queensland, St Lucia QLD 4072, Australia

**Email**: hongfu.sun@uq.edu.au

**Abbreviations used:** QSM, quantitative susceptibility mapping; ROI, region of interest; DGM, deep grey matte; CNN, convolutional neural network; OctConv, octave convolution; MEDI, Morphology Enabled Dipole Inversion; RESHARP, regularization enabled sophisticated harmonic artifact reduction for phase data; LN-QSM, least-norm QSM; COSMOS, calculation of susceptibility through multiple orientation sampling; STAR-QSM, streaking artifact reduction for quantitative susceptibility mapping; GP, globus pallidus; PU, putamen; CN, caudate nucleus; TH, thalamus; SN, substantia nigra; RN, red nucleus; RA, relative anisotropy; SNR, signal to noise ratio; PSNR, peak signal to noise ratio; SSIM, structural similarity; NRMSE, normalized root mean squared error; ppb, parts-per-billion; SSE, sum of squared errors; PS, patch size.

**ABSTRACT**

Quantitative susceptibility mapping (QSM) provides a valuable MRI contrast mechanism that has demonstrated broad clinical applications. However, the image reconstruction of QSM is challenging due to its ill-posed dipole inversion process. In this study, a new deep learning method for QSM reconstruction, namely xQSM, was designed by introducing noise regularization and modified octave convolutional layers into a U-net backbone and trained with synthetic and *in vivo* datasets, respectively. The xQSM method was compared with two recent deep learning (QSMnet$^+$ and DeepQSM) and two conventional dipole inversion (MEDI and iLSQR) methods, using both digital simulations and *in vivo* experiments. Reconstruction error metrics, including peak signal to noise ratio (PSNR), structural similarity (SSIM), normalized root mean squared error (NRMSE), and deep grey matter susceptibility measurements, were evaluated for comparison of different methods. The results showed that the proposed xQSM network trained with *in vivo* datasets achieved the best reconstructions among all deep learning methods. In particular, it led to, on average, 32.3%, 25.4%, and 11.7% improvement in the accuracy of globus pallidus susceptibility estimation for digital simulations and 39.3%, 21.8%, and 6.3% improvements for *in vivo* acquisitions, when comparing to DeepQSM, QSMnet$^+$, and iLSQR, respectively. It also exhibited the highest linearity against different susceptibility intensity scales and demonstrated the most robust generalization capability to various spatial resolutions among all deep learning methods. In addition, the xQSM method also substantially shortened the reconstruction time from minutes using MEDI to only a few seconds.

# INTRODUCTION

Quantitative susceptibility mapping (QSM) is an MRI post-processing technique, which extracts magnetic tissue susceptibility from MRI phase images[1,2]. It has shown great potential for studying a variety of neurological disorders, such as healthy aging[3], Multiple Sclerosis[4,5], Alzheimer's disease[6], Parkinson's disease[7], alcohol use disorders[8], and intracranial hemorrhage[9]. However, QSM reconstruction is non-trivial, involving several key post-processing steps[1,2]. For example, the raw gradient-echo phase images from multiple receivers need to be correctly combined[9] and unwrapped[10]. Background magnetic field contribution from sources outside of the brain should be removed[11-14]. Dipole inversion is then performed, which is an ill-posed inverse problem[1,2].

Different methods have been proposed to solve the ill-posed dipole inversion. One method uses multiple orientation sampling to compensate for the missing data in a single orientation, which is known as the Calculation of Susceptibility through Multiple Orientation Sampling (COSMOS)[15]. This method requires at least three different head orientations to solve the dipole deconvolution analytically. Although COMSOS is considered as the gold standard except for anisotropic white matter[16], the time-consuming repeated scans and the requirements for patients to rotate heads hinder its feasibility in the clinic. Therefore, iterative methods such as iLSQR[17], morphology enabled dipole inversion (MEDI)[18], truncated dipole inversion[19], and total field inversion methods[20-22] were developed to restore high-quality susceptibility maps from single orientation measurements. However, these methods can be prone to artifacts, computationally intensive, time-consuming, and requiring manual parameter tuning for different datasets.

Recently, deep learning has been applied to solve a variety of inverse problems[23] as an alternative to the conventional iterative methods, owing to its capability to approximate any continuous function, given enough learnable parameters and large datasets[23]. The hypothesis[23-25] is that compared with traditional methods using explicitly designed regularization, deep neural networks can learn more effective data-driven regularization to preserve anatomical structures more accurately. Lastly, deep learning-based image reconstructions are generally fast[23-25], which is a significant advantage in practice.

Previous studies have successfully trained multiple deep neural networks for QSM dipole inversion. QSMnet[26] trained a 3D U-net[27] with single-orientation local field and COSMOS maps as inputs and labels, and recently, this work was extended to QSMnet$^+$ [28] using data augmentation to improve the linearity of the original QSMnet. Alternatively, QSMGAN[29] improved this scheme by adding a discriminator on the U-net to construct a Generative Adversarial Network (GAN)[30]. Another latest deep learning work trained a variational network to perform non-linear dipole inversion (VaNDI)[31] using the same training dataset from QSMnet. However, these methods did not consider the susceptibility anisotropy effects when using COSMOS as the training labels. AutoQSM[32] proposed to train a U-net that could restore brain tissue QSM directly from the total field maps without explicit skull stripping and background field removal. The training inputs were single-orientation total field maps, while the training labels were QSM images reconstructed using the two-step STAR-QSM[33] pipeline, which preserved the susceptibility anisotropy. All of the above four frameworks (i.e., QSMnet/QSMnet$^+$, QSMGAN, VaNDI, and AutoQSM) require reconstructing QSM firstly using conventional iterative methods as the training labels; however, these assigned labels may not be the ground truth due to reconstruction errors. In other words, the training inputs and labels used in these studies do not strictly follow the dipole convolution physical model[11,22]. In contrast, another deep learning framework, DeepQSM[34], also based on the original U-net, generated local field maps from synthetic susceptibility labels of simple geometric shapes using the forward model so that the training inputs and labels satisfy the exact underpinning equation between susceptibility source and induced field. However, QSM results degraded noticeably when the test data deviate from the model (e.g., noise and error in measurements) in DeepQSM[34,35] and susceptibility underestimation was observed, especially in deep grey matter (DGM) regions.

Octave convolution (OctConv)[36] is a recently proposed operation, aiming to reduce the substantial redundancy that exists in the spatial dimension of feature maps generated in convolutional neural networks (CNNs). The OctConv operation explicitly factorizes the feature maps in CNNs into high and low spatial resolution groups, which corresponds to the fact that natural images can be decomposed into high- and low-frequency components. It has been reported[36] that this design can improve the efficiency of the networks and accomplish better multi-scale representation for various image and video tasks. In this study, we propose an enhanced U-net framework, namely xQSM, for QSM dipole inversion via incorporating modified OctConv layers, aiming to improve the accuracy and robustness of deep learning-based QSM reconstruction and recover the susceptibility contrast loss, particularly susceptibility underestimation in the DGM regions. Furthermore, a tailored noise adding layer is incorporated in training the network to regularize the noise and error amplification from the ill-posed dipole inversion for *in vivo* acquisitions. The effects of training datasets, noise levels, image resolutions, as well as susceptibility value ranges on the deep neural networks, is also thoroughly investigated across two different field strengths.

**METHODS**

**Ill-posed QSM Dipole Inversion**

The biological tissue with magnetic susceptibility distribution $\chi(r)$, where $r$ denotes the spatial location when placed in the main field $B_0$ of an MRI scanner, will gain a magnetization in the z-direction $M_Z(r) \approx \chi(r) \cdot B_0/\mu_0$, where $\mu_0$ is the vacuum permeability. This tissue magnetization will generate a magnetic field perturbation $\Delta B(r)$, which can be formulated as:

$$\Delta B(r) = \frac{\mu_0}{4\pi} \int d^3 r \{\frac{3(r-r')\cdot M(r')\cdot (r-r')}{|r-r'|^5} - \frac{M(r')}{|r-r'|^3}\}, \qquad (1)$$

and simplified as:

$$\Delta B(\boldsymbol{r}) = \chi(\boldsymbol{r}) \otimes d(\boldsymbol{r}) \cdot B_0, \qquad (2)$$

where $d(\boldsymbol{r}) = (3\cos^2\theta - 1)/(4\pi |\boldsymbol{r}|^3)$ is referred to as the unit dipole kernel, and $\theta$ is the angle between $\boldsymbol{r} - \boldsymbol{r}'$ and the main field $\boldsymbol{B}_0$; $\otimes$ represents the convolution operation. This convolution relation can be simplified as a multiplication in the k-space[11,22]:

$$\chi(\boldsymbol{k}) \cdot d(\boldsymbol{k}) = \Delta B(\boldsymbol{k})/B_0, \qquad (3)$$

where $d(\boldsymbol{k})=1/3-k_z^2/(k_x^2+k_y^2+k_z^2)$ (d = 0 at the origin) is the unit dipole kernel in k-space, and $\boldsymbol{k}=[k_x,k_y,k_z]$ are the k-space coordinates. Equation (3) depicts the ill-posed inverse problem with zero values of dipole kernel $d(\boldsymbol{k})$ on a double conical surface ($k_x^2+k_y^2=2k_z^2$). On the other hand, it also serves as a well-posed forward model for preparing the training datasets in this work.

**xQSM Deep Neural Network**

The proposed xQSM network is an enhanced 3D U-net with all conventional convolutions replaced with the new OctConv operations (detailed in the following section), as shown in Fig. 1. Similar to the original U-net, the xQSM network comprises two CNNs: a contracting and an expanding part. Such a design can improve the efficiency of the network and reduce the memory cost during training. The proposed xQSM also has two concatenations from layers of contracting parts to expanding parts, which compensate for the spatial information loss after max-pooling layers. This concatenation design can mitigate the gradient vanishing problem and improve network performance[25]. The proposed xQSM also adds a skip connection between the input and the output, forming a residual block[37], which helps the training process converge faster. This skip connection can further mitigate the vanishing gradient problem during training[38] and provide a noticeable performance enhancement compared to networks without residual connections[25]. Moreover, a noise adding layer (detailed in a later section) is inserted before the training inputs to regularize and suppress the noise and error amplification from the ill-posed dipole inversion. As summarized and illustrated in Fig. 1, the xQSM network contains 10 OctConv layers ($3^3$ kernel size with stride 1), 2 max-pooling layers ($2^3$ kernel size with stride 1), 2 transposed convolution layers ($2^3$ kernel size with stride 2), 12 batch normalization layers, 1 final convolution layer ($1^3$ kernel size with stride 1), and 1 noise adding layer. The rectified linear unit (ReLU) is adopted as the activation function of the network.

**Octave Convolution**

The original OctConv[36] was proposed to reduce the redundancy and improve the efficiency of deep neural networks. In this study, we replaced the nearest-neighbour interpolation in OctConv with a transposed convolution to allow for more learnable parameters. As illustrated in the middle row of Fig. 1, the new OctConv is formulated as a combination of 8 basic operations (i.e., four 3D traditional convolutions, one 3D average pooling, one 3D transposed convolution, and two additions):

$$\begin{aligned} Y_H &= F_{HH} + F_{LH} \\ Y_L &= F_{HL} + F_{LL} \\ F_{HH} &= \text{Conv}_{HH}(X_H) \\ F_{HL} &= \text{Conv}_{HL}(\text{AvgPool}(X_H)) \\ F_{LH} &= \text{ConvT}(\text{Conv}_{LH}(X_L)) \\ F_{LL} &= \text{Conv}_{LL}(X_L) \end{aligned} \quad (4)$$

where $\text{Conv}(\cdot)$ represents convolution operations with subscripts indicating different kernels (i.e., $H$ for High resolution and $L$ for Low resolution); $\text{ConvT}(\cdot)$ represents the conventional transposed convolution of kernel size 2, which doubles the resolution of feature maps; $\text{AvgPool}(\cdot)$ is the average pooling operation of stride 2, which halves the resolution of feature maps; $X \in \mathbb{R}^{h \times w \times d \times c}$ denotes the input feature maps in traditional convolution networks, where h, w, d, and c represent the height, width, depth, and channels of this feature, respectively. The OctConv explicitly factorizes the feature maps into two spatial resolution groups $\{X_H, X_L\}$ along the channel dimension, where $X_H \in \mathbb{R}^{h \times w \times d \times (\alpha_X)c}$ represents high spatial resolution group and $X_L \in \mathbb{R}^{h/2 \times w/2 \times d/2 \times (1-\alpha_X)c}$ is the low spatial resolution group, and $\alpha_X$ is a ratio factor. Similarly, the output features $Y \in \mathbb{R}^{h \times w \times d \times c}$ in traditional convolutions can also be decomposed into $\{Y_H, Y_L\}$ and serve as inputs to the next OctConv layer, and $\alpha_Y$ is the ratio factor for the factorization of the output feature maps. In this work, $\alpha_X$ and $\alpha_Y$ were set to 0.5 for input and output feature maps of all the middle layers. For the first layer, $\alpha_X$ (input) was set to 1 and $\alpha_Y$ (output) was set to 0.5 to convert the conventional feature maps into octave feature maps. For the last OctConv layer, $\alpha_X$ was set to 0.5 and $\alpha_Y$ was set to 1, resulting in a single high-resolution output[36]. Note that although the complexity of the OctConv layers is higher, each of the OctConv kernels (Conv$_{HH}$, Conv$_{HL}$, Conv$_{LH}$, Conv$_{LL}$) has only one-fourth learnable parameters of the traditional convolutional kernel, and thus the total number of learnable parameters are kept similar.

**Noise Adding Layer**

A noise adding layer (shown as the yellow arrow in Fig. 1), which adds different levels of Gaussian noise into the training inputs, was designed in this work to improve the robustness of the xQSM network against the noise and error amplification, which suppresses the streaking artifacts from the ill-posed dipole inversion. The operation of this noise adding layer is defined as:

$$\begin{aligned} Y &= X + I * Noise \\ I &= \begin{cases} 0, & \text{if } rand > P \\ 1, & \text{if } rand < P \end{cases} \\ Noise &= \sqrt{Power/SNR} * randn \\ Power &= (\sum X^2)/numel(X) \\ SNR &= randSNR([40, 20, 10, 5]) \end{aligned} \tag{5}$$

where $X$ and $Y$ are input maps before and after noise adding layer; $P$ is the probability of adding noise into the input patch; *rand* is a function that returns a uniformly distributed random number in the interval of (0, 1); *randn* generates a standard normal distribution matrix of the designated image size; *numel*($X$) returns the number of all elements in the image matrix $X$; and *randSNR* randomly returns one of the four input SNRs (i.e., 40, 20, 10, 5) with equal probability. These random functions will be activated from batch to batch during network training.

In this work, the probability parameter $P$ was set empirically to 20%, which means for each mini-batch during training, there is a 20% probability of adding noise of a certain level (i.e., randomly chosen from the SNR list [40, 20, 10, 5]) into the training inputs. This parameter setting also means that 80% iterations of the network training process were based on the QSM data without noise to reinforce the underlying physics model. An xQSM network without the noise adding layer was also trained to investigate the benefits of the proposed noise regularization.

**Data Preparation**

Local field maps, as training inputs, were generated via convolving the QSM labels with the unit dipole kernel according to the forward model in Eq. (3), which is consistent with the DeepQSM study[34]. Two sets of QSM labels were studied: one set from *in vivo* QSM brain volumes acquired from 90 healthy subjects (1 mm isotropic resolution) and reconstructed with the LN-QSM method[22], and another set from synthetic digital phantoms containing basic geometric shapes (spheres, squares, and rectangles) as described in the DeepQSM paper[34].

QSM datasets were cropped into small patches to fit into GPU memory for training. For *in vivo* dataset preparation, as shown in the top row of Fig. 1, the full-size (144×196×128) brain QSM volumes were cropped into small patches (size $48^3$) via firstly sliding a $48^3$ cropping window with a stride of 24×36×20 while traversing the 90 LN-QSM full volumes to obtain 11,250 patches, followed by randomly cropping 3,750 more patches from all 90 subjects (around 41-42 patches from each subject). After cropping, the corresponding local field maps of the same size (i.e., $48^3$) were computed according to the forward model (Eq. (3)). In total, 15,000 *in vivo* susceptibility patches (size $48^3$) and their corresponding field maps were generated, as labels and inputs for network training. For the synthetic dataset, we directly simulated 15,000 volumes of synthetic phantoms (size $48^3$) as described in the DeepQSM paper[34], and then calculated their corresponding field maps. The proposed xQSM deep neural networks trained on these two sets of QSM data (i.e., *in vivo* and synthetic) were referred to as $xQSM_{invivo}$ and $xQSM_{synthetic}$.

**Network Training**

The xQSM networks were optimized via minimizing the following L2-norm loss function:

$$\arg\min_{\theta} \frac{1}{2N} \sum_{i=1}^{N} \|C(X_i;\theta)-Y_i\|_F^2, \tag{6}$$

where $\{X_i, Y_i\}_{i=1}^{N}$ denotes the $N$ number training inputs and labels; $\|\cdot\|_F^2$ is the Frobenius norm; $\theta$ represents the hyper-parameters of the networks; and $C(X_i;\theta)$ is the output of the current deep neural network. In this study, each xQSM network was trained for around 18 hours (i.e., 100 epochs) using 2 Tesla V100 GPUs with a mini-batch size of 32. All weights and biases were initialized with normally distributed random numbers with a mean of zero and a standard deviation of 0.01. The xQSM network parameters were optimized using adaptive moment estimation (Adam optimizer[39]). The learning rate and was set to $10^{-3}$, $10^{-4}$, and $10^{-5}$ for the first 50 epochs, 50-80 epochs, and the final 20 epochs, respectively; all other hyper-parameters were set to their default values.

The proposed xQSM deep neural network was implemented using MATLAB R2019a. Although the network was trained with cropped patches of size $48^3$, the reconstruction implementation using xQSM network can operate on full-size local field maps without cropping. Indeed, the size of local field maps is required to be evenly dividable by 8, and zero-padding can be performed to satisfy this requirement. The source codes and trained networks ready for evaluation on new local field datasets are published at https://github.com/sunhongfu/deepMRI/tree/master/xQSM.

**Validation with Simulated and *in vivo* Datasets**

The proposed xQSM$_{invivo}$ and xQSM$_{synthetic}$ were compared with two conventional dipole inversion methods (i.e., iLSQR[17], and MEDI[18]) and another two deep learning methods (QSMnet$^+$[28] and DeepQSM[34]). The DeepQSM network presented here was trained with the same synthetic dataset used for xQSM$_{synthetic}$, while the QSMnet$^+$ model adopted in this work was downloaded from the original paper (https://github.com/SNU-LIST/QSMnet) and was not re-trained.

Different dipole inversion methods were first compared using digital simulations. (i)

A magnetic field map of the brain was generated by the forward model from a COSMOS map reconstructed from five head orientations (1 mm isotropic at 3T). (ii) Four local field maps of two contrast levels and two noise levels (denoted by Sim1Snr1, Sim1Snr2, Sim2Snr1, and Sim2Snr2, with "Sim" representing contrast level and "Snr" standing for noise level) from the 2019 QSM Challenge 2.0 (http://qsm.snu.ac.kr/?pageid=30) were also tested. (iii) A 3D Shepp-Logan digital phantom with various ellipsoids inside (susceptibilities of -100, -250, -200, -50, 150, and 350 parts-per-billion (ppb)) was constructed as shown in Fig. 5 to test the generalization capability of the deep learning-based QSM methods. Peak signal to noise ratio (PSNR), structural similarity (SSIM), and region of interest (ROI) measurements were evaluated.

An ultra-high-resolution COSMOS map of 0.6 mm isotropic voxel size acquired at 7T, was resized to (i) 1 mm isotropic, (ii) 2 mm isotropic, and (iii) $0.6 \times 0.6 \times 2$ mm$^3$ to study the effects of image resolution mismatch on deep learning-based QSM methods. Normalized root mean square errors (NRMSEs) were recorded for quantitative comparisons. The susceptibilities from ground truth data of 2019 QSM Challenge 2.0 were multiplied with different scaling factors (i.e., 0.3, 0.5, 1, 1.5, 2, 2.5, 3) to investigate the impacts of susceptibility intensity range on the deep learning reconstructions. Linear regressions were carried out to evaluate the linearity performance of different deep learning QSM methods.

For the *in vivo* experiments, the phase images from neutral head orientation acquisitions from the ten healthy subjects (five at 3T with 1 mm isotropic resolution and five at 7T with 0.6 mm isotropic resolution) were processed by the standard QSM pipeline, including phase unwrapping using the best path method [10] and background field removal using the RESHARP method [13]. Different dipole inversion methods were then performed on these pre-processed local field maps. Susceptibilities of globus pallidus (GP), putamen (PU), caudate nucleus (CN), thalamus (TH), substantia nigra (SN), and red nucleus (RN) were measured and compared with those from the gold standard COSMOS maps. Deep learning-based QSM methods were also performed on local field maps acquired with the toward-left-shoulder and toward-right-shoulder head orientations from a 1 mm COSMOS subject to examine the white matter susceptibility anisotropy effect[16].

The local field map processed with the standard QSM pipeline from a Multiple Sclerosis (MS) patient with multiple T1 blackhole lesions (1 mm isotropic at 3T) was also tested for evaluating the generalization capability of the proposed xQSM network, which was trained using healthy brains. Visual inspections were performed to identify and assess the delineation of white matter lesions in different QSM results.

**RESULTS**

**Effects of the Noise Adding Layer**

The xQSM results with and without the noise adding layer, on an *in vivo* local field map from a healthy volunteer, were compared with the COSMOS reconstruction in Fig. 2. It is clearly shown in the full axial slice and the zoomed-in frontal white matter region that $xQSM_{invivo}$ trained with the noise adding layer resulted in substantially suppressed noise amplification as compared to training without any noise regularization. The line profiles crossing the basal ganglia also confirmed that the proposed noise adding layer helped reduce the large oscillations of the susceptibility measures (red circles) and maintained the susceptibility contrast. However, the mean susceptibility measurement of the DGM remains unchanged with (e.g., GP 152 ± 30 ppb) and without noise adding layer (e.g., GP 153 ± 40 ppb).

**Digital Brain Simulation**

Reconstruction results of xQSM, $QSMnet^+$, DeepQSM, iLSQR, and MEDI for local field maps simulated from a COSMOS map (1 mm isotropic at 3T) were shown in Fig. 3. The proposed $xQSM_{invivo}$ showed similarly high PSNR and SSIM (44.95/0.97) as the conventional MEDI method (46.07/0.96). The $xQSM_{synthetic}$, $QSMnet^+$, and DeepQSM substantially underestimated susceptibilities of DGM (yellow arrows), especially GP with 19.8%, 24.3%, and 44.6% underestimation, respectively. Comparisons of deep learning-based methods (Fig. 3 bottom table) suggested that the performance of DeepQSM (i.e., U-net trained with synthetic datasets) can be improved by incorporating the OctConv operation as in xQSM$_{synthetic}$, and resulted in 5% ~ 24.8% of improvements in DGM regions. In addition, the network trained with synthetic datasets (xQSM$_{synthetic}$) can be further enhanced with *in vivo* training data (xQSM$_{invivo}$) with an improvement of 1.6 ~ 22.6% in DGM regions. For this dataset (size 192×256×176), the reconstruction time of deep learning methods (about 2.4 seconds for xQSM, 1.4 seconds for DeepQSM, and 2 seconds for QSMnet$^+$) was substantially shorter than traditional methods (about 48 and 614 seconds for iLSQR and MEDI, respectively).

**2019 QSM Challenge Datasets**

Figure 4 compared the results of different deep learning and conventional QSM methods on the 2019 QSM Challenge 2.0 data (1 mm isotropic). The proposed xQSM$_{invivo}$ led to the best PSNR and SSIM among all deep learning methods and comparable reconstruction results to iLSQR and MEDI. As shown in the error maps, xQSM$_{synthetic}$, QSMnet$^+$, DeepQSM, and iLSQR methods displayed susceptibility underestimation in DGM regions, especially in GP (17.6%, 28.7%, 22.2%, and 12.0% underestimation, respectively, as reported in Fig. 4 bottom table). In comparison, xQSM$_{invivo}$ and MEDI did not show noticeable underestimation in the DGM region (on average 8.0% underestimation in all six regions for xQSM$_{invivio}$ and 3.8% for MEDI). It was confirmed that the susceptibility underestimation in DGM regions introduced by DeepQSM was alleviated by incorporating the proposed OctConv (e.g., 4% improvement from DeepQSM to xQSM$_{synthetic}$ in GP), or by training with *in vivo* datasets (17% improvement from xQSM$_{synthetic}$ to xQSM$_{invivo}$) or implementing both strategies (21% improvement from DeepQSM to xQSM$_{invivo}$). The reconstruction speeds of deep learning methods (about 2.2 seconds for xQSM, 1.4 seconds for DeepQSM, and 2.2 seconds for QSMnet$^+$) were considerably faster than iLSQR and MEDI (54 and 720 seconds) for all the Challenge datasets (image size 164×205×105).

**Digital Shepp-Logan Phantom**

Figure 5 compares deep learning QSM reconstructions on a digital 3D Shepp-Logan phantom for testing the generalization capability of these trained neural networks. The susceptibility and error maps from different methods showed that QSMnet$^+$ introduced the most susceptibility contrast loss, while the xQSM$_{synthetic}$ produced the most accurate susceptibility map (PSNR/SSIM: 36.65/0.97) among all deep learning methods. The error metrics suggested that the xQSM networks trained on synthetic data, containing simple shapes only, performed better than *in vivo* brain datasets on the Shepp-Logan phantom. Susceptibility measurements of all the uniform structures are displayed in the bar graph of Fig. 5, which confirmed that the proposed xQSM$_{synthetic}$ was the most accurate method followed by DeepQSM and xQSM$_{invivo}$.

**Generalization Evaluation of the xQSM method**

Deep learning-based QSM results from different spatial resolutions are demonstrated in Fig. 6. Both the error maps and the numerical metrics (i.e., NRMSE) indicated that the proposed xQSM$_{invivo}$ performs the best for all image resolutions, including isotropic and anisotropic voxel sizes. For isotropic resolutions, the performance of all methods degrades when the resolution is higher than that of the training data, while improves when the resolution goes lower. All methods produced the most substantial errors in the anisotropic resolution case, with NRMSE ranging from 22.1% to 47.3%. However, xQSM$_{invivo}$ is substantially more robust than the other deep learning methods, and DeepQSM is the most prone to input and training image resolution mismatch. The minimum GP susceptibility variation across different image resolutions was achieved using xQSM$_{invivo}$ (198 ± 15 ppb) as compared to xQSM$_{synthetic}$ (162 ± 34 ppb), QSMnet$^+$ (151 ± 54 ppb), and DeepQSM (125 ± 62 ppb). Evident DGM susceptibility underestimation is observed in the error maps from xQSM$_{synthetic}$, QSMnet$^+$, and DeepQSM, which is significantly reduced in the xQSM$_{invivo}$ results. The bar graph in Fig. 6 reporting the GP measurements from different methods also confirmed that the xQSM$_{invivo}$ led to the best GP measurements (less than 1% error) for all spatial resolutions tested.

Figure 7 illustrates the effect of different susceptibility intensity ranges on the deep learning QSM reconstructions. Figure 7(a) showed similar trends for all deep learning methods that NRMSE decreases with susceptibility scaling factors. The xQSM$_{invivo}$ exhibited the smallest NRMSE and appeared flattest across all scaling factors. Globus pallidus susceptibility measurement graph in Fig. 7(b) showed the highest accuracy and the best linearity with xQSM$_{invivo}$ (slope: 97, sum of squared error (SSE): 38).

**In vivo Experiments**

The average maps of COSMOS and other single-orientation QSM methods from ten *in vivo* subjects (five 0.6 mm isotropic at 7T, five 1 mm isotropic at 3T) were shown in Fig. 8. The zoomed-in DGM regions showed that xQSM$_{synthetic}$ and DeepQSM from 1 mm as well as xQSM$_{synthetic}$, QSMnet+, and DeepQSM from 0.6 mm substantially underestimated (pointed by the yellow arrows) DGM susceptibility compared with the COSMOS maps. It is also noted that QSMnet$^+$ and DeepQSM performed noticeably worse in 0.6 mm than 1 mm case, while xQSM$_{synthetic}$ and particularly xQSM$_{invivo}$ were more robust in the 0.6 mm results. DGM susceptibility measurements from different methods are evaluated against COSMOS in the bottom row of Fig. 8, for both resolutions. Paired t-tests found significant DGM susceptibility underestimation from xQSM$_{synthetic}$ (14.6%, $P = 0.0051$), DeepQSM (26.5%, $P = 0.0005$), and iLSQR (12.3%, $P = 0.0096$) in the 1 mm case. For the 0.6 mm case, xQSM$_{invivo}$ resulted in 21.2% ($P = 0.0013$), 37.1% ($P = 0.0002$), and 50.8% ($P = 0.0001$) accuracy improvement in GP susceptibility than xQSM$_{synthetic}$, QSMnet$^+$, and DeepQSM, respectively.

Deep learning QSM results on two *in vivo* local field maps (1 mm isotropic at 3T) from two different head orientations (i.e., toward-left-shoulder and toward-right-shoulder) are shown in Fig. 9. After registering QSM results from both orientations to the neutral head position, susceptibilities of the left and right internal capsule from different deep learning-based QSM methods were summarized in the table at the bottom of Fig. 9. A relative anisotropy (RA) of the measurements from two head orientations (e.g., a and b) was defined as |a-b|/|a+b| for evaluating the susceptibility anisotropy from different methods. As shown in the table, xQSM$_{invivo}$ (on average 36.5% RA), xQSM$_{synthetic}$ (on average 57% RA), and DeepQSM (on average 75% RA) showed more substantial orientation-dependent magnetic susceptibility anisotropy in internal capsule (red arrows) than QSMnet$^+$ (on average 16% RA). Even though QSMnet$^+$ showed the least susceptibility anisotropy, it can suppress the artifacts more than the other methods (e.g., the dark susceptibility artifact in the frontal area).

The proposed xQSM and other dipole inversion methods were applied to a MS subject acquired at 3T with 1 mm isotropic resolution to validate the generalization capability of the deep learning-based QSM method in Fig. 10. As shown in the first column, the T1-weighted images displayed multiple MS lesions (red arrows), appearing as "black holes". All QSM methods successfully detected the T1 lesions as indicated by the red arrows (bottom row). However, the DeepQSM method showed reduced susceptibility contrast (top row), and the MEDI reconstruction showed severe artifacts that may obscure the lesions (bottom row).

**DISCUSSION**

In this work, we developed a new deep learning framework – xQSM – to perform the ill-posed dipole inversion of QSM reconstruction. The original convolutional layers in a U-net, as used in other deep learning QSM methods[26,28,32,34], were replaced by the modified OctConv[36] layers, which explicitly factorizes the feature maps into high- and low- resolution groups and introduces inter-group operations to allow communication of information between different groups. Such factorization improves the capability of neural networks for multi-scale representation learning[36], which is beneficial for patch-based QSM network training. We also designed a noise adding layer to regularize the training of the proposed xQSM network and improve its capability to suppress amplification from potential noise and measurement deviations in the *in vivo* field maps. Our xQSM results showed that the DGM susceptibility underestimation present in U-net based methods was successfully recovered by the modified OctConv layers, and the noise adding layer design substantially suppressed the artifacts and enhanced signal-to-noise ratio in the QSM results.

Previous works have proposed to solve QSM dipole inversion with deep neural networks, including QSMnet[26], QSMnet[+28], QSMGAN[29], VaNDI[31], autoQSM[32], and DeepQSM[34]. The first five frameworks require reconstructing QSM first using conventional methods as the training labels, which may deviate from the ground truth. DeepQSM[34] proposed to learn the underpinning physics of dipole inversion with synthetic data. The DeepQSM scheme has two main advantages: (i) the training input and the output satisfy the physical relation between magnetic field and susceptibility source, and (ii) randomly generated synthetic data eliminates the requirement for large datasets of *in vivo* QSM data acquisitions for network training. However, susceptibility underestimation in DGM was reported in the DeepQSM method. Our xQSM results showed that the susceptibility underestimation in GP region from DeepQSM (nearly 36% underestimation on average) was alleviated via incorporating the proposed modified OctConv (around 18% underestimation) or combining the OctConv with *in vivo* training datasets (less than 3% underestimation).

The improvement from U-nets to xQSM networks may originate from the feature factorization design in OctConv. In traditional convolutional layers, all feature maps are in the same spatial dimension (i.e., image size). However, some of the feature maps may represent low-frequency information, which means that the spatial resolution of these feature maps can be compressed to reduce the spatial redundancy and enhance the training efficiency as in OctConv. The OctConv has been tested in various network backbones and showed impressive improvements over traditional convolutions[36]. With the OctCov factorization, our xQSM network indeed consists of two sub-U-nets of different resolutions with frequent information exchange from layer to layer. The low-resolution U-net convolutes more spatial information (i.e., broader dipole coverage) than the high-resolution U-net, and this effectively enlarges the receptive fields[40], which is beneficial for the modelling of the non-local convolutional relation between the magnetic field and susceptibility source. The enhancement of the proposed factorization design is also consistent with a previous study[41], suggesting that multiple branches improve the performance of deep networks.

Various metrics including PSNR, SSIM, and DGM measurements, were compared between xQSM, QSMnet$^+$, DeepQSM, iLSQR, and MEDI methods. The comparison was carried out on both simulated and *in vivo* datasets, including COSMOS-simulated field maps, 2019 QSM Challenge 2.0 datasets, a simulated Shepp-Logan phantom, and thirteen *in vivo* field maps. The results illustrated in simulated and *in vivo* datasets suggest that the proposed xQSM$_{invivo}$ achieved the best dipole inversion among all deep learning methods. Besides, compared with traditional iterative methods, the reconstruction speed of deep learning based methods is an essential and practical advantage. For example, the reconstruction of the proposed xQSM is only 4.2 seconds compared to 129 seconds of iLSQR and over 1000 seconds of MEDI on an image of size 224×304×224.

The proposed xQSM achieved the most robust dipole inversion results against a variety of susceptibility intensity ranges and image spatial resolutions, which indicates that xQSM has the best generalization capability among all deep learning QSM methods evaluated in this work. For example, xQSM$_{invivo}$ is the only deep learning method that produced comparable image contrast to COSMOS at the 0.6 mm resolution without substantial susceptibility contrast loss. This robustness to the image resolution may originate from the OctConv design, which explicitly factorizes the feature maps into high- and low-resolution groups. For the implementation, these two groups are feature maps of different resolutions, and via this design, a multi-scale signal representation is constructed in each layer of the proposed xQSM, which increases the network's generalization capability as compared to traditional convolutional layers[36].

Besides, the proposed xQSM networks, along with DeepQSM, successfully preserved the white matter susceptibility anisotropy, which is absent in COSMOS-trained networks such as QSMnet$^+$ due to their isotropic susceptibility assumption.

Five different training patch sizes ($16^3$, $32^3$, $48^3$, $64^3$, and $80^3$) were investigated in this work, with PSNR, SSIM, NRMSE, and the DGM susceptibility measurements compared. The effect of training patch size on the proposed xQSM$_{invivo}$ is shown in Supp. Fig. 1. It is observed that the xQSM network trained with the smallest patch size (PS) of $16^3$ led to significant errors in the high-susceptibility DGM region. All networks achieved similar reconstruction results when the PS was equal to or larger than $32^3$. The DGM susceptibility measurements are reported in Supp. Table 1, with the most accurate susceptibility measurements produced with a PS of 48, followed by 64. Larger patch sizes led to limited improvement when evaluating the PSNR, SSIM, and NRMSE in Supp. Table 2, at the cost of significantly increased training time and GPU memory requirement. For example, it takes about 20 hours to train the network with a patch size of $48^3$, requiring two Tesla V100 GPUs, while it takes over 40 hours to train the network based on patches of $80^3$, requiring four Tesla V100 GPUs due to memory demand. Therefore, by balancing reconstruction accuracy and computing load, this study chose a patch size of $48^3$ to train the xQSM networks, which is consistent with QSMnet[26] and QSMGAN[29].

Even though the proposed networks were trained with cropped patches of size $48^3$, the reconstruction implementation using xQSM network can operate on full-size local field maps without cropping, owing to the invariance of the convolutional operation to different image sizes. In this work, all xQSM results were conducted via the "full-size" manner, i.e., taking the full-size local field maps as the network inputs. Alternatively, the trained xQSM model can also be applied for QSM reconstruction through a "patch-then-assemble" manner if the GPU memory is too small to directly reconstruct a full-size image. The full-size images were first cropped into small patches by sliding a cropping window to traverse the full-size local field maps. These small patches were then fed into the pre-trained xQSM network to obtain the corresponding patch-based QSM reconstructions. Finally, these patched QSM results were assembled to obtain full-size QSM reconstructions. The xQSM reconstructions on a simulated COSMOS subject (1 mm isotropic resolution) using three different patch sizes (i.e., $16^3$, $32^3$, and $48^3$) with a fixed stride ($4^3$) are compared in Supp. Fig. 2. It can be seen that although the full-size reconstruction achieved the best result, the "patch-then-assemble" approach can lead to acceptable QSM reconstructions, as long as the PS is equal to or larger than $32^3$.

To improve the ill-posed dipole inversion robustness over noise amplification and streaking artifacts, we designed a noise adding layer to address this problem. It is shown that this noise adding layer design resulted in xQSM$_{invivo}$ with significantly enhanced signal-to-noise ratio while maintained the susceptibility contrast and accuracy when compared with COSMOS. However, in the current model, the noise added into the network training was normally distributed, and the magnitude of the noise is limited to four pre-defined levels, which may deviate from the real-world situations. For example, the proposed xQSM method slightly underperformed the MEDI method using the 2019 QSM Challenge data (Fig. 4) due to the noise distribution and level mismatch.

The network trained with *in vivo* brain datasets (i.e., xQSM$_{invivo}$) achieves better reconstruction on the brain datasets, while the network trained with synthetic geometrical shapes (i.e., xQSM$_{synthetic}$) performs better on the Shepp-Logan phantom (Fig. 5). This impact of training datasets on QSM deep neural networks has also been reported in a recent deep learning QSM review paper[35]. To further investigate the training dataset impact, another hybrid network (xQSM$_{Hybrid}$) was trained with a mix of synthetic and *in vivo* datasets. Reconstruction results of xQSM$_{invivo}$, xQSM$_{synthetic}$, and xQSM$_{Hybrid}$ for a local field map simulated from a COSMOS label (1 mm isotropic) were compared. It was found that the xQSM$_{invivo}$ network performed the best among the three different training datasets, demonstrating the minimum error map (Supp. Fig. 3), the highest PSNR and SSIM, and the lowest NRMSE ( Supp.Table 3).

In the future, more realistic and comprehensive noise distributions, as well as more comprehensive designs of the training datasets, will be thoroughly studied and employed for the noise adding layer and training of the network, respectively, to make the framework more robust against MRI data of various noise levels and achieve more accurate QSM dipole inversion results.

**CONCLUSION**

We proposed a new deep learning framework- xQSM, for fast and robust QSM dipole inversion. By incorporating octave convolution and noise regularization layers and training with *in vivo* brain images through the forward modeling, xQSM method suppresses streaking artifacts, alleviates noise amplification, enhances robustness to different resolution and image intensity ranges, and significantly reduces the deep grey matter susceptibility underestimation in previous deep learning-based QSM methods.

**Acknowledgment**

We thank Steffen Bollmann and Markus Barth from the University of Queensland for sharing their codes for generating the synthetic training dataset in the present work. We also acknowledge the Queensland Brain Institute for providing high-performance GPU computing support. H.S. acknowledges grant support from the University of Queensland (UQECR2057605) and the Australian Research Council (DE210101297). G.B.P. acknowledges grant support from the Canadian Institutes of Health Research (FDN-143290) and the Natural Sciences and Engineering Research Council of Canada (DG-03880). For collection of the 7T human imaging data the authors acknowledge the facilities and scientific and technical assistance of the National Imaging Facility, a National Collaborative Research Infrastructure Strategy (NCRIS) capability, at the Melbourne Brain Centre Imaging Unit, The University of Melbourne. Participant informed consent and Human Research Ethics Committee approval was obtained for collection of the 3T and 7T data.

**FIGURE LEGENDS**

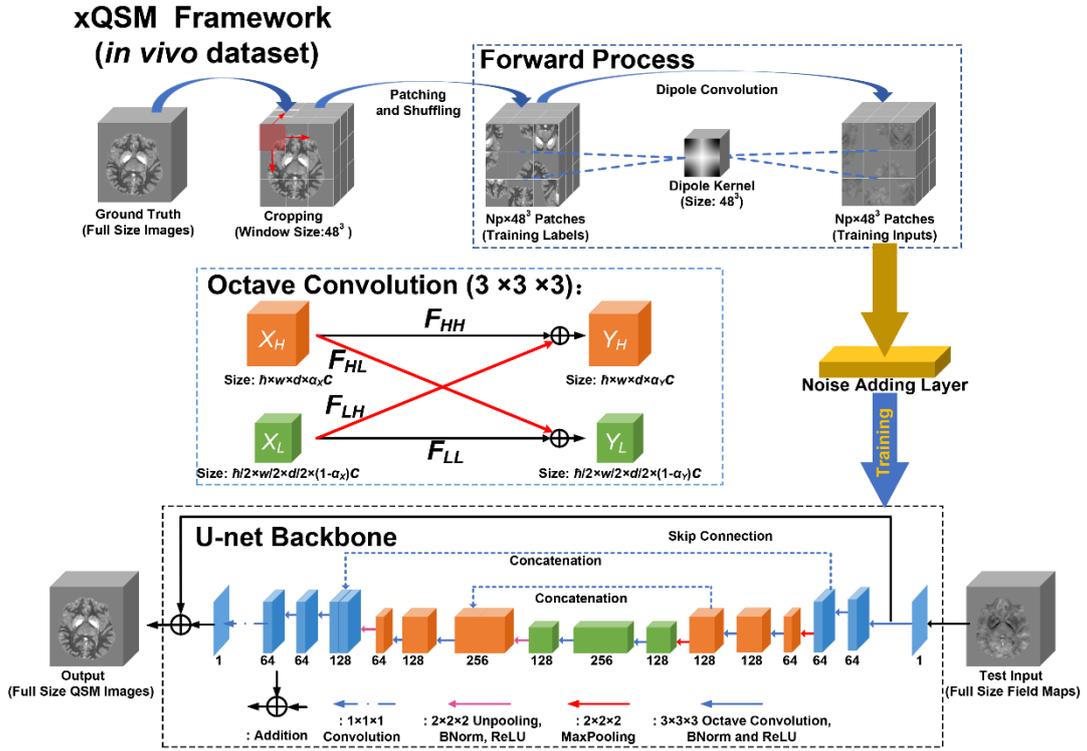

**Figure 1:** Overview of the proposed xQSM method. The top row demonstrates the preparation process with the *in vivo* training datasets. Octave convolution is shown in the middle row, which introduces an X-shaped operation for communication between feature maps of different resolutions. Training input patches pass through a noise adding layer (yellow) during each iteration step. The bottom row illustrates the xQSM network architecture based on the U-net backbone.

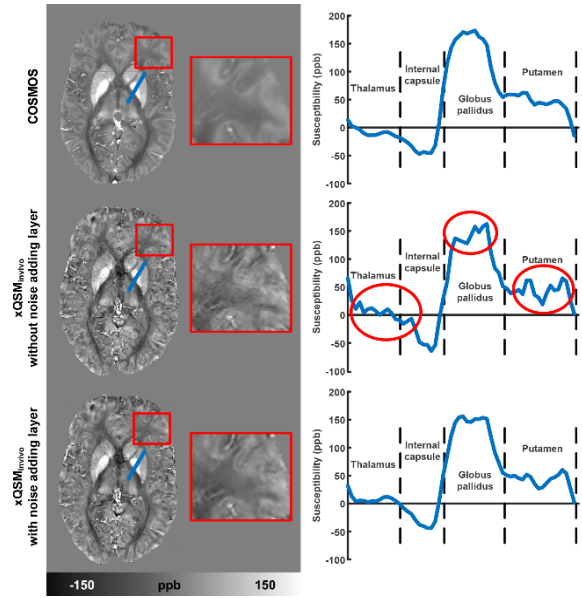

**Figure 2:** Comparison of the proposed xQSM$_{invivo}$ networks trained with and without the noise adding layer on one *in vivo* local field map. Axial mid-brain slice containing DGM and zoomed-in frontal white matter region are displayed in the left column. The line profiles crossing DGM and internal capsule are plotted in the right column, with red circles highlighting the noisy and oscillating susceptibility measurements.

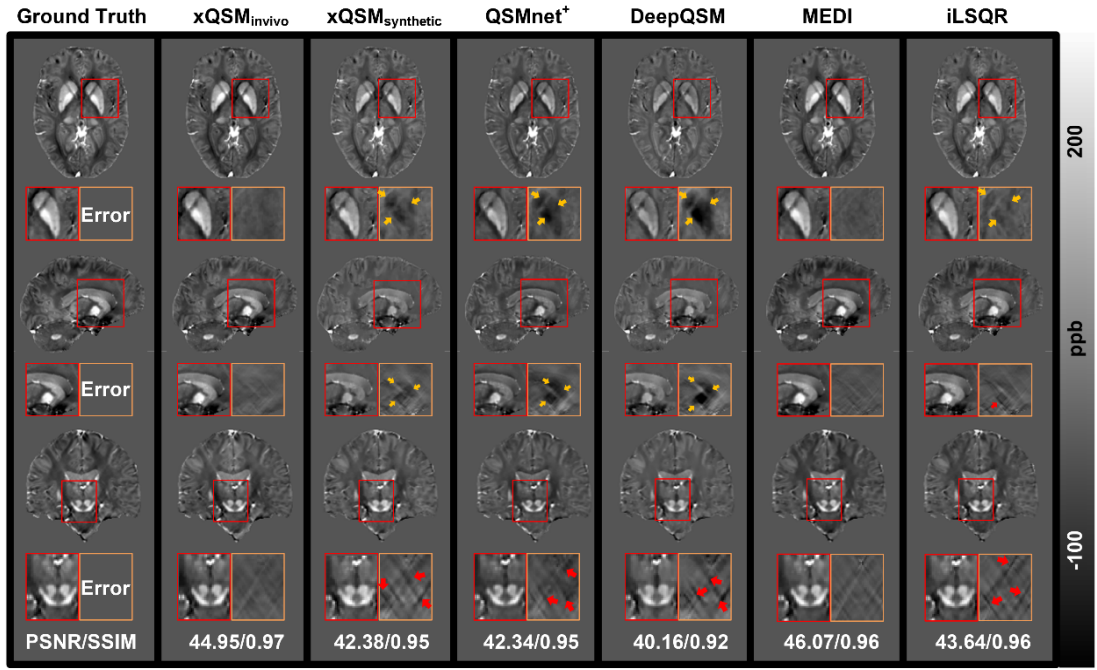

**Figure 3:** Comparison of different QSM methods on a COSMOS (1 mm isotropic) simulated data in three orthogonal views. Yellow arrows point to significant DGM susceptibility underestimation, while red arrows indicate apparent streaking artifacts in the error maps. PSNR and SSIM relative to the ground truth are reported under the images in white font. DGM susceptibility measurements and percentage errors from different methods are summarized in the table at the bottom.

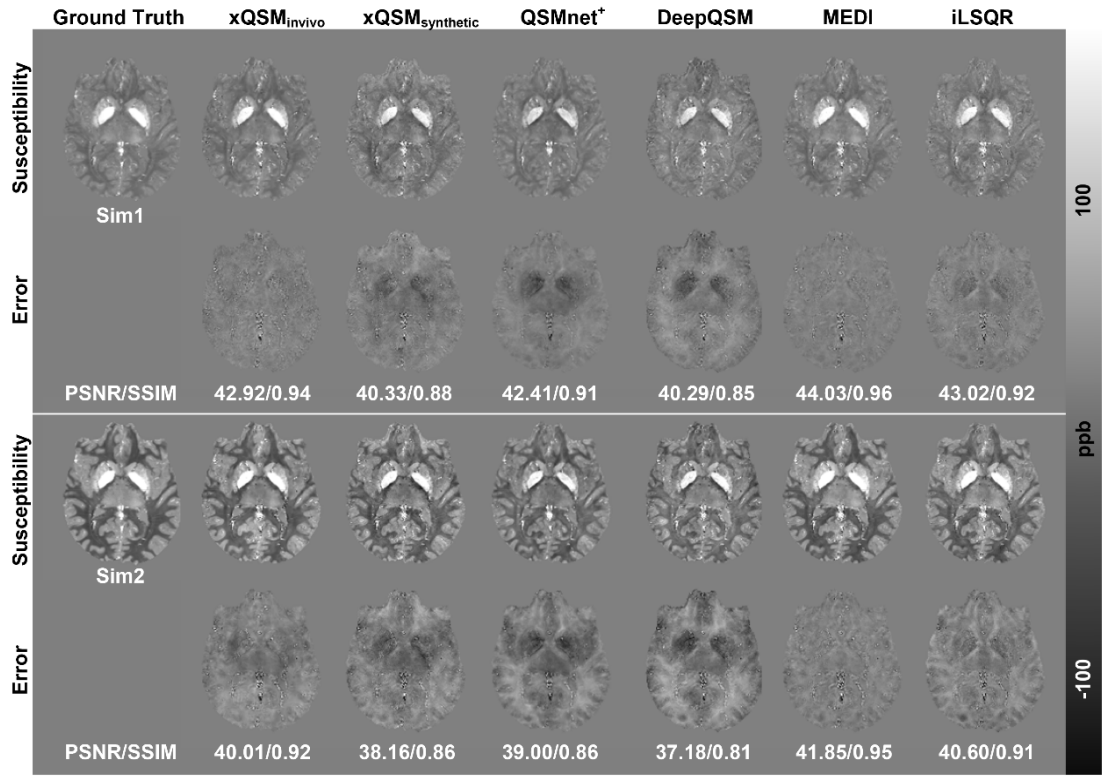

**Figure 4:** Comparison of different QSM methods on the 2019 QSM Challenge data. The top two rows illustrate the reconstruction and error maps of the lower contrast level data (the average of "Sim1Snr1" and "Sim1Snr2"), while the bottom two rows of the high-contrast data (the average of "Sim2Snr1" and "Sim2Snr2"). The table at the bottom reports the DGM susceptibility mean and standard deviation measurements for each method.

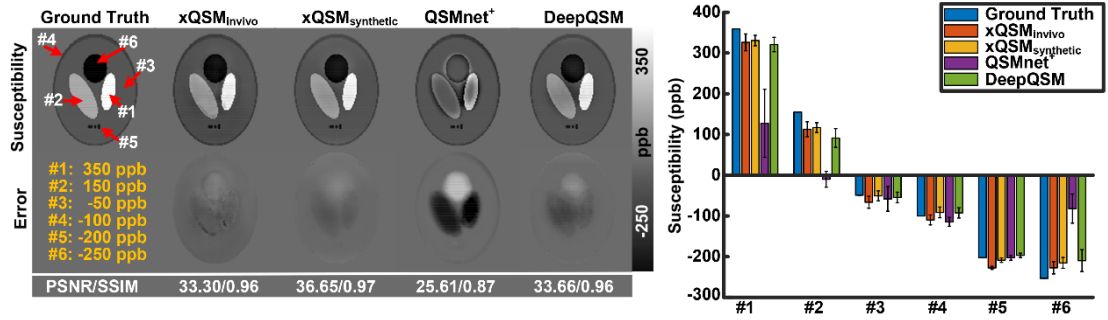

**Figure 5:** Comparison of different deep learning methods on a 3D Shepp-Logan phantom. QSM results and error maps are presented along with PSNR and SSIM. The bar graph reports the corresponding ROI measurements (mean and standard deviation) of the six regions identified by the red arrows.

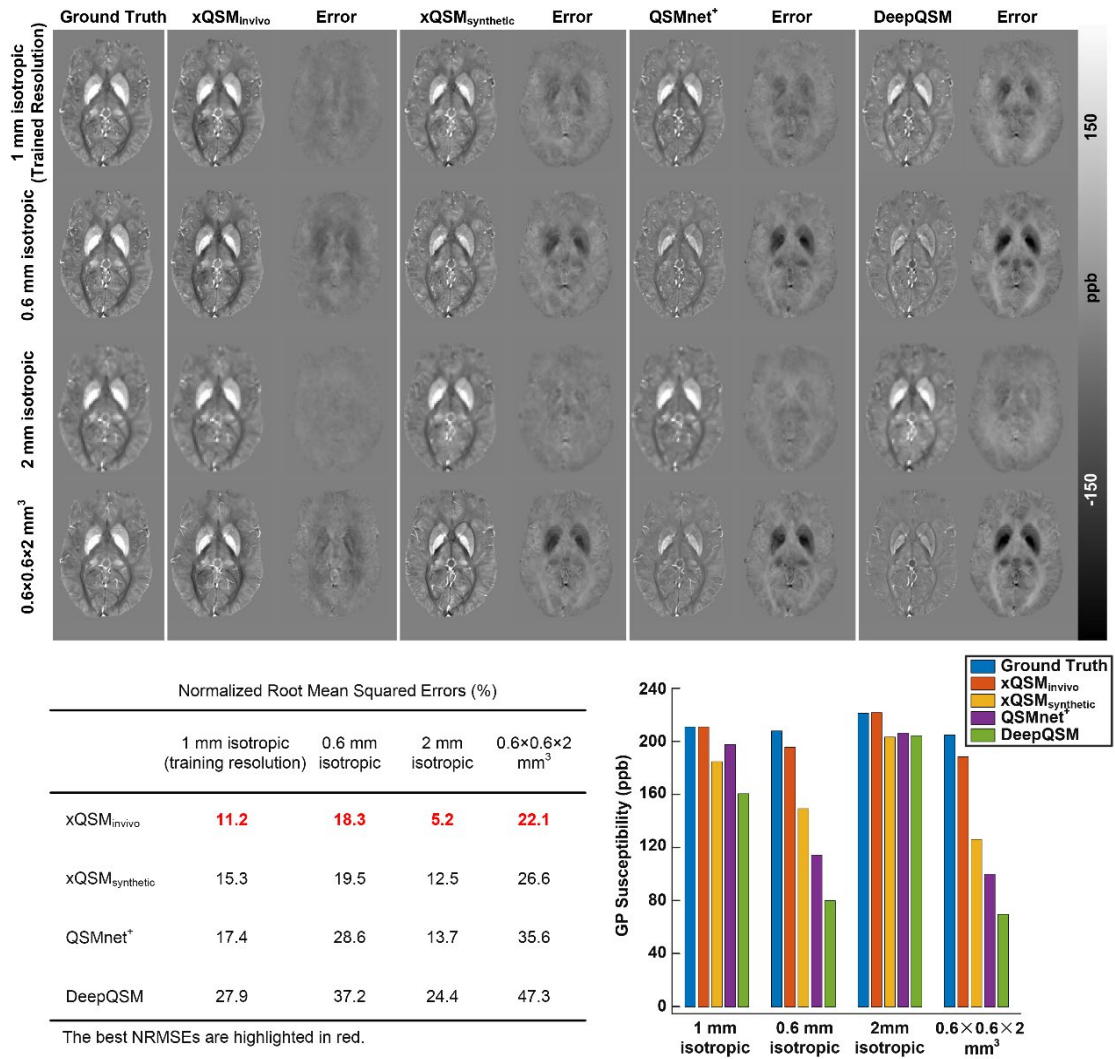

**Figure 6:** QSM results and error maps (relative to ground truths) from four deep neural networks are compared on various spatial resolutions in the top four rows.

NRMSE for each method and spatial resolution is reported in the bottom table. The globus pallidus (GP) susceptibility measurements from different methods at different spatial resolutions are compared in the bar graph.

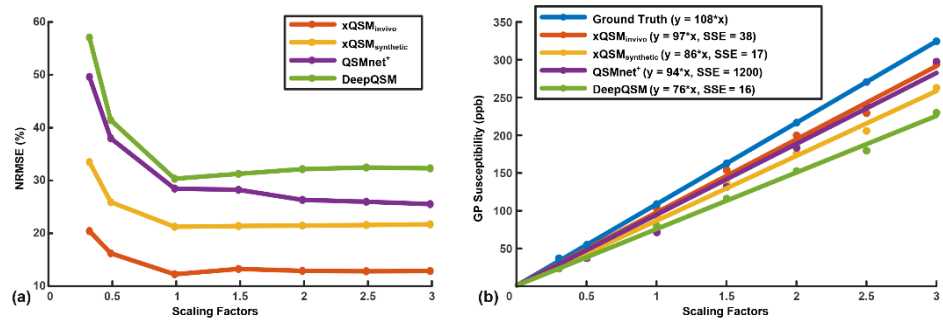

**Figure 7:** Linearity performance comparison of different deep learning-based QSM methods against varying susceptibility ranges, using the 2019 QSM Challenge data. The line graph on the left plots the trend of the NRMSE (%) relative to the ground truth data as the scaling factors increase. The right line graph illustrates the linear regression of GP measurements against susceptibility intensity scaling factors for each method. Slopes and SSEs are reported in the legend of (b).

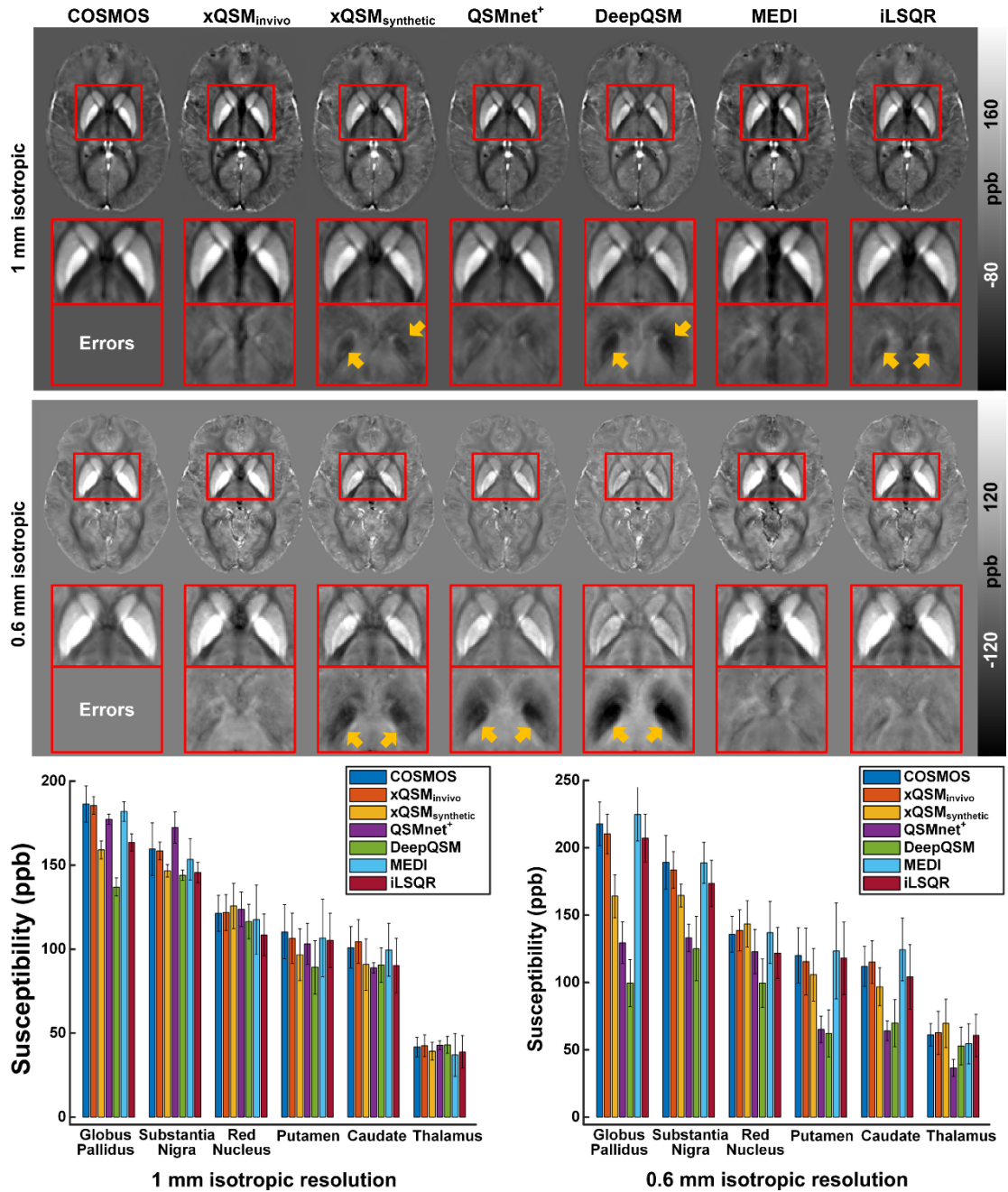

**Figure 8:** Comparison of different QSM methods on ten *in vivo* local field maps (five 0.6 mm isotropic from 7T and five 1 mm isotropic from 3T). Average QSM maps and DGM zoomed-in images are shown in the top four rows. Yellow arrows point to the apparent DGM susceptibility contrast loss with respect to the COSMOS. The average DGM susceptibility measurements (mean and standard deviation) from all subjects are plotted at the bottom for 1 mm and 0.6 mm acquisitions, respectively.

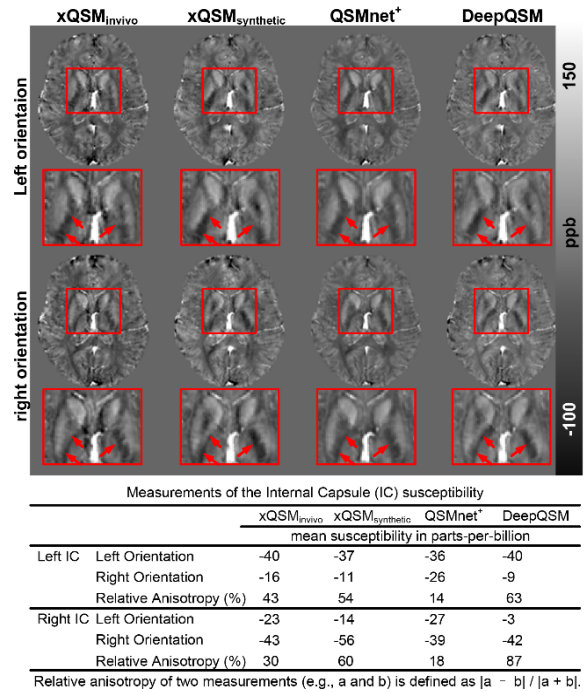

**Figure 9:** Deep learning QSM results on two *in vivo* local field maps (1 mm isotropic) acquired at two different head positions (i.e., toward-left-shoulder and toward-right-shoulder). Red arrows point to the head orientation-dependent susceptibility anisotropy in the region of the internal capsule. Susceptibilities of left and right internal capsule from the two head orientations are reported in the table.

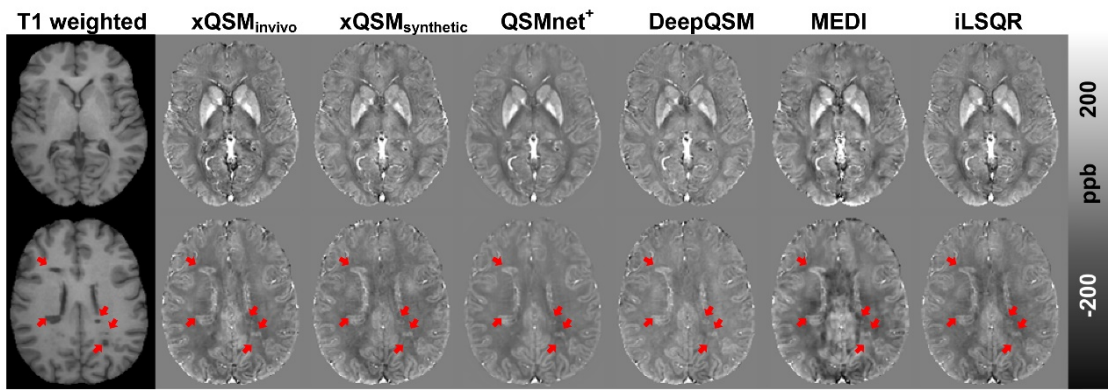

**Figure 10:** Different QSM methods on an *in vivo* field map (1 mm isotropic) acquired from a patient with Multiple Sclerosis. The first column shows T1-weighted magnitude images in two axial slices, and the registered QSM volumes from different methods are displayed starting the second column. Red arrows point to the MS lesions that are identified by visual inspection.